\documentclass[twocolumn,showpacs,showkeys,pra,floatfix]{revtex4}	
\usepackage{amsfonts}
\usepackage{amsthm}
\usepackage{amssymb}
\usepackage{amsmath}

\usepackage[dvips]{graphics}
\usepackage{epsfig}
\usepackage{euscript}
\usepackage{color}

\newtheorem{theorem}{Theorem}

\newcommand{\be}{\begin{equation}}
\newcommand{\ee}{\end{equation}}
\newcommand{\bea}{\begin{eqnarray}}
\newcommand{\eea}{\end{eqnarray}}
\newcommand{\beann}{\begin{eqnarray*}}
\newcommand{\eeann}{\end{eqnarray*}}

\newcommand{\ket}[1]{\vert{#1}\rangle}

\newcommand{\unity}{{1\hskip -3pt \rm{I}}}

\newcommand{\ip}[2]{\langle{#1}|{#2}\rangle}
\newcommand{\Hil}{\mathcal{H}}

\newcommand{\Spin}{J}

\newcommand{\J}{J}
\renewcommand{\a}{\alpha}

\begin{document}

\title{Implementing Quantum Gates using the Ferromagnetic Spin-$\Spin$
  XXZ Chain with Kink Boundary Conditions}

\author{Tom Michoel}
\email{tom.michoel@psb.vib-ugent.be}
\affiliation{Department of Plant Systems Biology, VIB}
\affiliation{Department of Molecular Genetics, Ghent University, Technologiepark 927, B-9052 Gent, Belgium.\\}

\author{Jaideep Mulherkar}
\email{jmulherkar@math.ucdavis.edu}

\author{Bruno Nachtergaele}
\email{bxn@math.ucdavis.edu}
\affiliation{Department of Mathematics, University of California, 
Davis, CA 95616-8633, USA.\\}

\begin{abstract}
We demonstrate an implementation scheme for constructing quantum gates using unitary evolutions of the one-dimensional spin-$J$ ferromagnetic XXZ chain.
We present numerical results based on simulations of the chain using the time-dependent DMRG method and techniques from optimal control theory.
Using only a few control parameters, we find that it is possible to implement one- and two-qubit gates on a system of spin-3/2 XXZ chains, 
such as Not, Hadamard, Pi-8, Phase, and C-Not, with 
fidelity levels exceeding $99\%$.
\end{abstract}



\maketitle

\section{Introduction}

For quantum computers to become a reality we need to find or build physical systems that faithfully implement the quantum 
gates used in the algorithms of quantum computation. The basic requirement is that the experimenter has access to two states
of a quantum system that can be effectively decoupled from environmental noise for a sufficiently long time, and that 
transitions between these two states can be controlled to simulate a number of elementary quantum gates (unitary transformations).
Systems that have been investigated intensively are atomic levels in ion traps \cite{CZ1995,MKIW1995}, superconducting device physics
using Josephson rings \cite{MOL1999}, nuclear spins \cite{CVZ1998}(using NMR in suitable molecules) and quantum dots 
\cite{LD1998}. In this paper we demonstrate  the implementation of quantum gates using one-dimensional spin-$J$ systems. The results are obtained using
a computer simulation of these systems.

The Hamiltonian of the XXZ model with kink boundary conditions is given by
\begin{align}
\label{eqn:Hamiltonian}
H_L^{\rm k}(\Delta^{-1})= &\sum_{\alpha=-L+1}^{L-1} \Big[(\Spin^2-S_\alpha^3 S_{\alpha+1}^3) - \Delta^{-1}(S_\alpha^1 S_{\alpha+1}^1\\ \nonumber
		 									      &+ S_\alpha^2 S_{\alpha+1}^2)\Big] + \Spin \sqrt{1-\Delta^{-2}}(S_{-L+1}^3 - S_L^3)\nonumber
\end{align}
where $S_\alpha^1$, $S_\alpha^2$ and $S_\alpha^3$ are the spin-$\Spin$ matrices acting on
the site $\alpha$. Apart from the magnitude of the spins, $J$, the main parameter of the model 
is the anisotropy $\Delta>1$ and the limit $\Delta\rightarrow\infty$ is referred to as the
\emph{Ising limit}. In the case of $J=1/2$ kink boundary conditions were first introduced 
in \cite{PS1990}. They lead to ground states with a domain wall between down spins on the left portion
of the chain and up spins on the right. The third component of the magnetization, $M$, is conserved,
and there is exactly one ground state for each value of $M$. Different values of $M$ correspond to 
different positions of the domain walls, which in one dimension are sometimes referred to as kinks.
In \cite{KNS2001}, Koma, Nachtergaele, and Starr showed that there is
a spectral gap above each of the ground states in this model for all values of $\Spin$. Recently \cite{MNSS2008}
it was shown that for spin values $J\ge \frac{3}{2}$ and for sufficiently large value of the anisotropy $\Delta$ 
the low lying spectrum of (\ref{eqn:Hamiltonian}) for each value of $M$ has isolated eigenvalues that persist in the thermodynamic limit.

The presence of isolated eigenvalues is ideal from the point of view of quantum computation.
The idea is to use the subspace, denoted by $\mathcal{D}$, of the ground state and the first excited state 
of the Hamiltonian to encode a qubit. We let the system evolve under its own unitary time evolution
generated by the Hamiltonian (\ref{eqn:Hamiltonian}) with the addition of a few local control fields. 
We have two requirements to fulfill: the time evolution should leave the qubit space $\mathcal{D}$ approximately
invariant, and the (approximately) unitary matrix describing the dynamics restricted to $\mathcal{D}$ and stopped 
at a suitable time should coincide with the desired quantum gate.
 
\begin{figure}
\begin{center}
\epsfig{file=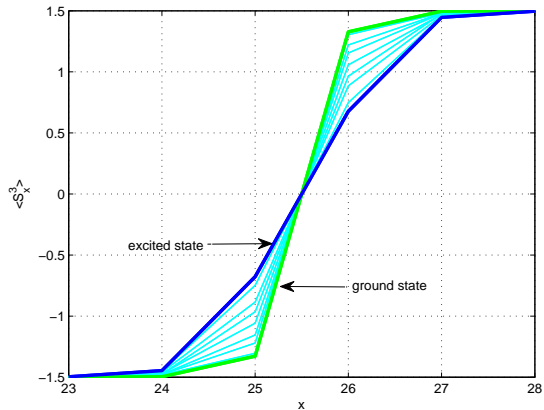,height=6cm,width=8cm}
\caption{\label{fig:proftrans}
The transitioning of the magnetization profile from the ground to the first excited state using a Not gate.
Simulation obtained for a chain of 50 sites using DMRG.}
\end{center}
\end{figure}

The control inputs needed to drive the system such that high fidelity 
gates are obtained are determined using techniques from optimal control theory. The simulation of the time evolution of the chain that is large enough to 
resemble the properties in the thermodynamic limit is carried out using the  Density Matrix Renormalization Group (DMRG) algorithm.
Figure \ref{fig:proftrans} shows the transition of the magnetic profiles in the z-direction from the ground to the first excited state using the Not gate constructed
from a spin-$\frac{3}{2}$ XXZ spin chain of length 50 sites. We also demonstrate the construction of Pi-8, Hadamard, and Phase gates that form a set of universal single qubit gates.

In order to have a viable quantum computing scheme one needs to implement at least one 2-qubit gate.
Here we have implemented the C-Not gate which, in combination with the 1-qubit gates, is known to be universal
\cite{AB1995}.

Our scheme capitalizes on the kink nature of the excitations of the XXZ Hamiltonian, which are rather sharply localized.
We imagine a setup with two parallel chains with the location of the kink lined up in their
ground states. The subspace for the 2-qubit state space 
is then $\mathcal{D}_1\otimes\mathcal{D}_2$, where $\mathcal{D}_1$ represents the space of isolated eigenvalues 
of the first chain and  $\mathcal{D}_2$ for the second chain. A set of three controls localized near the kinks is
used to generate the single qubit gates acting on  $\mathcal{D}_1$ and  $\mathcal{D}_2$ and a C-Not gate on 
$\mathcal{D}_1\otimes\mathcal{D}_2$. This scheme produces a universal set of gates necessary for two-qubit computation. 
It is clear how to generalize this scheme to implement n-qubit computation.
Since a universal set of single qubit gates and nearest neighbor C-Not gates are universal for n-qubit computation, 
this can be achieved by using $n$ parallel chains and controls that are localized and act on neighboring chains only. 

In the next section we describe the model and review some of the past results.
Then, in section \ref{sec:Control}, the optimal control problem to construct the quantum gates is described.
Section \ref{sec:DMRG} is devoted to the DMRG algorithm and the specific 
adaptations to the XXZ spin chain. Finally, in section \ref{sec:Results} we present our results based on
numerical simulations of the XXZ Hamiltonian using the DMRG algorithm.

\section{The Model}
\label{sec:Model}
In this section we describe in detail the spin-$J$ ferromagnetic XXZ model with kink boundary conditions on the one-dimensional lattice 
$\mathbb{Z}$. The local Hilbert space for a single site $\alpha$ is $\mathcal{H}_\alpha = \mathbb{C}^{2\J +1}$ with $\Spin \in \frac{1}{2} \mathbb{N} = \{0,\frac{1}{2},1,\frac{3}{2},2,\dots\}$.
We consider the Hilbert space for a finite chain on the sites $[-L+1,L] = \{-L+1,-L+2,\dots,+L\}$.
This is $\mathcal{H}_{[-L+1,L]}=\bigotimes_{\alpha=-L+1}^L\mathcal{H}_\alpha$. 
The Hamiltonian of the spin-$\Spin$ XXZ model is given by equation (\ref{eqn:Hamiltonian}).
Note that, by a telescoping sum, we can absorb the boundary fields
into the local interactions:
\begin{eqnarray*}
\label{XXZ Hamiltonian}
H_L^{\rm k}(\Delta^{-1})		 &=& \sum_{\alpha=-L+1}^{L-1} h^{\rm k}_{\alpha,\alpha+1}(\Delta^{-1})\\
h^{\rm k}_{\alpha,\alpha+1}(\Delta^{-1}) &=&  \Spin^2-S_\alpha^3 S_{\alpha+1}^3 - \Delta^{-1}(S_\alpha^1 S_{\alpha+1}^1 +  S_\alpha^2 S_{\alpha+1}^2)\\
					 &+& \Spin \sqrt{1-\Delta^{-2}}\, (S_{\a}^3 - S_{\a+1}^3)
\end{eqnarray*}
The main parameter of the model is the anisotropy $\Delta>1$ and we get the Ising limit as $\Delta\rightarrow\infty$. 
It is mathematically more convenient to work with the parameter $\Delta^{-1}$, which we then assume is in the interval $[0,1]$.
As we said, $\Delta^{-1}=0$ is the Ising limit, and $\Delta^{-1}=1$ is the isotropic XXX Heisenberg model.
The Hamiltonian commutes with the total magnetization
$$
S^3_{\rm tot}\, =\, \sum_{\alpha =-L}^L S^3_\alpha\, .
$$
As indicated in the introduction, for each $M \in \{-2\Spin L,-2\Spin L +1 ,\dots,2\Spin L\}$,
the corresponding sector is defined to be the eigenspace of $S^3_{\rm tot}$ with eigenvalue $M$; clearly, these are invariant 
subspaces for all the Hamiltonians introduced above. These subspaces are called ``sectors''.

It was shown \cite{PS1990,ASW1995,GW1995,KN1998} that the kink boundary conditions
lead to a family of ground states.
It was also shown in \cite{ASW1995,GW1995,Mat1996,KN1998} that for each sector 
there is a unique ground state of $H_L^{\rm k}(\Delta^{-1})$ with eigenvalue 0. Moreover, this ground state, $\psi_M$, is given
by the following expression:
\begin{equation*}
\psi_M = \sum
\bigotimes_{\alpha \in [-L+1,L]}\binom{2\Spin}{\Spin -m_\alpha}^{1/2} q^{\alpha(\Spin-m_\alpha)}\ket{m_\alpha}_\alpha\, ,
\end{equation*}
where the sum is over all configurations for which $\sum_\alpha m_\alpha = M$
and the relationship between $\Delta>1$ and $q\in(0,1)$ is given by $
\Delta=(q+q^{-1})/2$.
A straightforward calculation shows a sharp transition in the magnetization from fully polarized down at the left to fully polarized up at the right.
For this reason they are called kink ground states.  In \cite{KNS2001}, Koma, Nachtergaele, and Starr showed that there is
a spectral gap above each of the ground states in this model for all values of $\Spin$. Recently \cite{MNSS2008} we were able to prove the following theorem.

\begin{theorem}
For spin values $J \ge 3/2$, there exists a finite $\Delta_0$ so that for all $\Delta > \Delta_0$, the first few excitations of $H_L^{\rm k}(\Delta^{-1})$
when restricted to any sector of magnetization, are isolated eigenvalues that persist in the thermodynamic limit. 
\end{theorem} 
In this paper it was also proved that in certain values of spin and sector, for example $\Spin = \frac{3}{2}$ and $M=0$ both the ground and excited states are non-degenerate (simple eigenvalues).
This is the qubit space we work with and our quantum gates will be unitaries on this space.

\section{Quantum gates using quantum control}
\label{sec:Control}
The problem of constructing quantum gates can be formulated as a problem in
quantum control theory \cite{MK2005}. The goal is to steer the system using a small number of 
control parameters such that the unitary operator describing the quantum dynamics after a finite time $T$,
has  maximal overlap with a desired target unitary (the gate). 
>From a  control perspective these problems reduce to control
of bilinear systems evolving on finite dimensional Lie groups. This is an optimal control problem on a 
two-level system which has been studied widely with exact results known in some cases. 
For example, time optimal implementation of single and two qubit quantum gates was studied \cite{KBG2001} 
when the Lie algebra $\rm{g}$ of $su(2)$ ($su(4)$) can be decomposed as a Cartan pair 
$\rm{g} = \rm {k} \oplus {p}$  with $ \rm{k}$ is the Lie subalgebra generated by a the drift Hamiltonian
and $\rm{p}$ is the Lie sub algebra generated by the control Hamiltonian's.
Finding the time optimal trajectories is reduced to finding geodesics on the coset space $G/K$ ($G$ and $K$ being the Lie Groups corresponding to $\rm{g}$ and \rm {k}).
The problem of driving the evolution operator while minimizing an energy-type quadratic cost was studied in \cite{DD2001}. In this case the optimal solutions
can be expressed as Elliptic functions. The time optimal problem of population transfer problem of a two-level quantum system and bounded controls
was studied in \cite{BM2005} and again explicit expressions for the optimal trajectories. In this paper we follow a numerical gradient based approach
to optimal control \cite{DD2008,KR2005}. 

\subsection{Single qubit gates}
\label{single qubit gates}
We consider the problem of time evolution of the one-dimensional XXZ chain under external controls.
The equation of motion for the unitary evolution of the XXZ chain isolated from the environment is given by Schr\"odinger's equation
\begin{equation}
\label{eqn:control1}
\dot{U}(t) = -i\Big(H_L^{\rm k}(\Delta^{-1}) + v(t)H^{\rm ext}\Big)U,\quad U(0) = \unity
\end{equation}
In control terminology $H_L^{\rm k}(\Delta^{-1})$ is the free or drift Hamiltonian and $H^{\rm ext}$ is the control Hamiltonian corresponding 
to the control field $v(t)$. We require that $\mathcal{D}$ is an invariant subspace of $H^{\rm ext}$, so that the time evolution of the system \ref{eqn:control1} 
given by the unitary $U(t)$ starting from an initial state in $\mathcal{D}$ will be constrained to $\mathcal{D}$ at all future times. The induced evolution on $\mathcal{D}$ 
at any specified final time $T$ will be the quantum gate on the qubit space $\mathcal{D}$ and is given by the 2x2 matrix

\begin{figure}
\epsfig{file=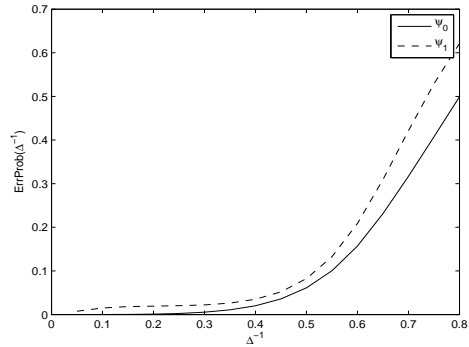,height=5cm,width=7cm}
\caption{
ErrProb, defined in (\ref{errprob}), as a function of $\Delta^{-1}$, 
calculated for the ground and first excited state $\psi_0$ and $\psi_1$,
with $H^{\rm ext}=S_0^3S_1^3$. This quantity provides a measure of the
escape rate out of the qubit subspace.}
\label{fig:err_prob}
\end{figure}

\begin{equation}
(U_{xxz})_{ij}:= \ip{\psi_i}{U(T)|\psi_j}\qquad i=0,1
\end{equation}
The control Hamiltonian we choose is the two site operator $H^{\rm ext}=S_0^3S_1^3$. In practice for $S_0^3S_1^3$  there is a very small error probability
for states to move out of $\mathcal{D}$ and the matrix $U_{xxz}$ is not exactly unitary. The matrix elements $\ip{\psi_0 | H^{\rm ext}}{\psi_k}$ and $\ip{\psi_1 | H^{\rm ext}}{\psi_k}$ $k \ne 0,1$  are proportional
to the transition probabilities to move from states $\psi_0$ and $\psi_1$ to other eigenstates of $H_L^{\rm k}(\Delta^{-1})$. We calculate the error probability to move out of the subspace $\mathcal{D}$ by the following
estimates of these matrix elements 
\begin{equation}\label{errprob}
\text{ErrProb} = \|H^{\rm ext}\psi_i\|^2 - |\ip{\psi_0}{H^{\rm ext}\psi_i}|^2 - |\ip{\psi_1}{H^{\rm ext}\psi_i}|^2 
\end{equation}
for $i=0,1$.
 Figure ~\ref{fig:err_prob} shows that the probabilities of transitioning out of the subspace $\mathcal{D}$ are extremely small for $\Delta^{-1} \le 0.3$. 

\subsection{Implementing two-qubit gates}
\label{two qubit gates}
The idea for implementing two-qubit gates is to use two copies of the XXZ chain. The Hilbert space for two-qubit quantum computation is 
$\mathcal{D}_1\otimes\mathcal{D}_2 \cong \mathbb {C}^4$, where $\mathcal{D}_1$ and $\mathcal{D}_2$ are the subspaces spanned by the ground state
and first excited state of the first chain and second chain respectively. The Hamiltonian of an uncoupled  two chain system is given by
\begin{equation*}
H_{L}^{\rm k}(\Delta^{-1})^{(1,2)} := H_L^{\rm k}(\Delta^{-1})^{(1)} +  H_L^{\rm k}(\Delta^{-1})^{(2)}
\end{equation*}
Here the notation  $H_L^{\rm k}(\Delta^{-1})^{(1)}$ is to be interpreted as $\underbrace{H_L^{\rm k}(\Delta^{-1})}_{chain 1}\otimes (\underbrace{\unity\otimes\cdots\otimes\unity}_{chain 2})$
and $H_L^{\rm k}(\Delta^{-1})^{(2)}$ is to be interpreted as $(\underbrace{\unity\otimes\cdots\otimes\unity}_{chain 1})\otimes \underbrace{H_L^{\rm k}(\Delta^{-1})}_{chain 2}$.
The two-qubit space is spanned by the four vectors $\psi_{mn}:=\psi_m\otimes\psi_n$ for $m,n=0,1$ which are eigenvectors of the above Hamiltonian.
If we consider the control system
\begin{align}
\label{eqn:control2}
\qquad\dot{U} =& -i\Big(H_{L}^{\rm k}(\Delta^{-1})^{(1,2)} + v_1(t) (S_0^3.S_1^3)^{(1)}\\ \nonumber
	        &+ v_2(t) (S_0^3.S_1^3)^{(2)}\Big)\nonumber
\end{align}
with  $U(0) = \unity$,
then by selectively turning on $v_1(t)$ and $v_2(t)$ for certain time periods, the above system is equivalent to the control system (\ref{eqn:control1}) on chains 1 and 2 respectively during those time intervals.
This can be used to generate single qubit gates on $\mathcal{D}_1$ and $\mathcal{D}_2$. Moreover by simultaneously using  $v_1(t)$ and $v_2(t)$ the local gates i.e. gates of the
kind $X_1\otimes Y_2$ can be generated on $\mathcal{D}_1\otimes\mathcal{D}_2$. To implement a two-qubit quantum computing scheme we need to also implement
perfectly entangling gates  i.e. a gate that can take a product state to a maximally entangled state. It is known that single qubit gates and any
perfectly entangling gate are universal for two-qubit quantum computing \cite{ZVSW2003}. Clearly such a gate cannot be implemented by the control scheme (\ref{eqn:control2}) alone.
In this paper we choose to implement the C-Not gate, which is an example of a perfectly entangling gate. For this purpose we make use of  an additional control namely  $(S_0^3 S_1^3)^{(1)}\otimes (S_0^3 S_1^3)^{(2)}$.

\begin{figure}
\begin{center}
\epsfig{file=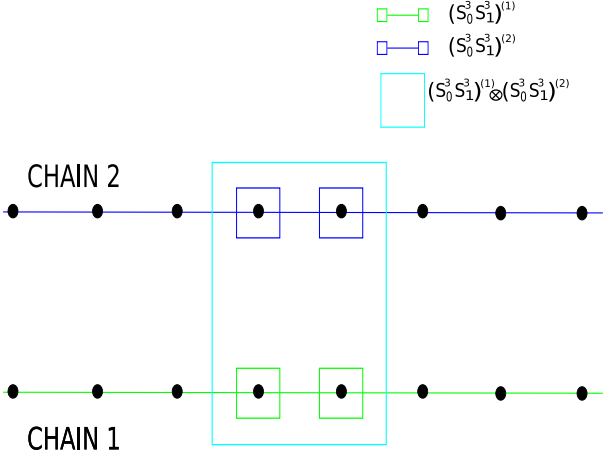,height=6cm,width=8cm}
\caption{\label{fig:xxz_schematic}
Configuration of two XXZ chains showing the localized controls 
required to implement the C-Not gate.}
\end{center}
\end{figure}

We  demonstrate the C-Not gate to high precision by using following control system   
\begin{align}
\label{eqn:control3}
\dot{U} = &-i\Big(H_{L}^{\rm k}(\Delta^{-1})^{(1,2)} + v_1(t) (S_0^3S_1^3)^{(1)} \\\nonumber
          +& v_2(t) (S_0^3S_1^3)^{(2)} + v_3(t)(S_0^3S_1^3)^{(1)}\otimes (S_0^3S_1^3)^{(2)}\Big)\nonumber
\end{align}
with $U(0) = \unity$ by selectively turning on and off some or all of the control fields $v_1(t)$, $v_2(t)$ and $v_3(t)$ for specified time periods. 
Figure \ref{fig:xxz_schematic} shows a diagrammatic representation of the two-qubit scheme. 
The C-Not gate is then given by the $4\times 4$ matrix with elements
\begin{equation*}
({\rm C-Not}^{xxz})_{mn;rs}:= \ip{\psi_{mn}}{U(T)|\psi_{rs}}\, i=0,1
\end{equation*}

\subsection{Optimal control}
\label{sec:optimal control}
We first solve the control problems (\ref{eqn:control1}) and (\ref{eqn:control3}) for the projected system  on $\mathcal{D}$ for the single chain and $\mathcal{D}_1\otimes\mathcal{D}_2$ for two chain system.
\begin{equation}
\label{eqn:control4}
\dot{U}(t) = -i\Big(H + \sum_k v_k(t)B_k\Big)U,\qquad U(0) = \unity
\end{equation}
For the projected system on $\mathcal{D}$ the $H$ and $B_k$'s are given by the $2\times 2$ matrices
\begin{eqnarray}
\label{eqn:matrix2}
H_{ij} &=& \ip{\psi_i}{H_L^{\rm k}(\Delta^{-1})|\psi_j}\\ \nonumber
(B_1)_{ij} &=& \ip{\psi_i}{H_L^{\rm k}(\Delta^{-1})|\psi_j}\, i,j=0,1 \nonumber
\end{eqnarray}
whereas the the projected system on $\mathcal{D}_1\otimes\mathcal{D}_2$ the control problem involves $4\times 4$ matrices
\begin{eqnarray}
\label{eqn:matrix4}
H_{mn;rs}&=& \ip{\psi_{mn}}{H_{L}^{\rm k}(\Delta^{-1})^{(1,2)}|\psi_{rs}}\\ \nonumber
(B_1)_{mn;rs}&=& \ip{\psi_{mn}}{(S_0^3.S_1^3)^{(1)}|\psi_{rs}}\\ \nonumber
(B_2)_{mn;rs}&=& \ip{\psi_{mn}}{(S_0^3.S_1^3)^{(2)}|\psi_{rs}}\\ \nonumber
(B_3)_{mn;rs}&=& \ip{\psi_{mn}}{(S_0^3.S_1^3)^{(1)}\otimes (S_0^3.S_1^3)^{(2)}|\psi_{rs}}  \nonumber
\end{eqnarray}
where $m,n,r,s =0,1$. The overlap between a desired unitary gate $U_f$ and the solution of (\ref{eqn:control3}) at time $T$, $U(T)$, is 
measured as the difference in the norm square $\|U_f-U(T)\|^2$, and the norm is defined in terms of the standard inner product $\ip{V}{W} := Tr(V^{\dag}W)$.
The norm  can be written as
\begin{equation*}
\|U_f-U(T)\|^2 = \|U_f\|^2 -2Re\ip{U_f}{U(T)} + \|U(T)\|^2
\end{equation*}
and hence minimizing this norm is equivalent to maximizing 
\begin{equation}
\label{eqn:cost}
\Phi:= Re\ip{U_f}{U(T)} = Tr(U_f^{\dagger}U(T))
\end{equation}
We define the gate fidelity as
\begin{equation}
\label{eqn:fidelity}
\mathcal{F}_{\text{Gate}}:= \frac{|Tr(U_f^{\dag}{U(T))}|}{Tr(\unity)}
\end{equation}
To select the optimal control fields $v_i(t)$ we use the numerical gradient ascent approach described in many books on control theory. This approach was applied to the
quantum setting in \cite{KR2005}. We start with the necessary conditions for optimality called the Pontryagin maximum principle which is a generalization of the Euler-Lagrange equations
from calculus of variations. In the problems with costs of type (\ref{eqn:cost}) and no a priori bound on controls, Pontryagin's maximum principle takes the following form 
\begin{theorem}(Pontryagin maximum principle \text{\cite{KR2005,BL2007}})
If $v_i(t)$'s are optimal controls of the system (\ref{eqn:control3}) and $U(t)$ the corresponding trajectory solution, then there exists a nonzero operator valued Lagrange multiplier $\lambda$ 
which is the solution of the adjoint equations 
\begin{eqnarray*}
\dot{\lambda}(t) &=& -iH(t)\lambda(t) \qquad \text{with terminal condition} \\
\lambda^{'}(T) &=&-\frac{\partial{\Phi(T)}}{\partial{U(T)}} = -U_f
\end{eqnarray*}
and a scalar valued Hamiltonian function $h(U(t),v_i(t)):= Re\ Tr(-i \lambda^{'}(t)H(t)U(t))$ such that, for every $\tau\in (0,T]$ we have
\begin{equation}
\label{eqn:gradient}
\frac{\partial {h(U)}}{\partial{v_i}}= Im\ Tr(\lambda^{'}(t) B_iU(t)) = 0
\end{equation}
\end{theorem}
The algorithm to find the optimal controls is as follows
\begin{enumerate}
\item
A suitable gate time $T$ is chosen and discretized in $N$ equal steps of duration $\Delta t = \frac{T}{N}$. 
The initial control $v_i^{(0)}(t_k)$ for all the discretized time intervals is based on a guess or at random.
\item
For these piecewise constant controls, from $U(0) = \unity$ and $\lambda(T)= -U_f$, compute the forward and backward propagation 
respectively as follows
\begin{eqnarray}
\label{eqn:fequation}
U^{(r)}(t_k)      &=& F^{(r)}(t_k)F^{(r)}(t_{k-1})\ldots F^{(r)}(t_1)\\ 
\label{eqn:bequation}
\lambda^{(r)}(t_k)&=& F^{(r)}(t_k)F^{(r)}(t_{k+1})\ldots F^{(r)}(t_N)\lambda(T) 
\end{eqnarray}
for all $t_1,\ldots,t_N$ and where $r$ is an iteration number of the algorithm initially set to 0 and
\begin{equation*}
\label{eqn:evolution1}
F^{(r)}(t_k) = exp\Big\{-i\Delta t\Big(H + \sum_i v_i^{(r)}(t_k)B_i\Big)\Big\}
\end{equation*}

\item
Substitute the equations (\ref{eqn:fequation}) and (\ref{eqn:bequation}) into equation (\ref{eqn:gradient}) to evaluate the gradient, and then update the controls
as $$v_i^{(r+1)}(t_k) = v_i^{(r)}(t_k) + \tau \frac{\partial h(U(t_k),v_i(t_k))}{\partial v_i}$$
where $\tau$ is a small step size.
\item if $\mathcal{F}_{\text{Gate}}< \gamma$ ($\gamma$ being the level of accuracy) then done, otherwise goto step (2) for the next iteration with the updated controls.
\end{enumerate}
Having solved the control problem on the projected systems to get the optimal controls $v_1(t)$, $v_2(t)$ and $v_3(t)$ we would like to 
apply them to a large system and see their effects on the projected system. However simulating even a moderately sized spin chain is hard because of
the exponentially growing dimension of the Hilbert space. In the next section we describe an algorithm by which we are able to simulate the XXZ chain of 50 sites.

\section{DMRG simulations for quantum gates}
\label{sec:DMRG}

To see the effect of the evolution of the XXZ chain with external magnetic controls we numerically simulate the XXZ chain using the DMRG algorithm.
The dynamics of the interfaces of the XXZ chain using DMRG was studied recently in \cite{MNS2007}.
The standard DMRG algorithm is a numerical algorithm originally developed by Steven White \cite{SW1993} that has worked successfully in providing very accurate results for ground state energies and 
correlation functions in strongly correlated systems. 
Modifications to this method \cite{GV2004,WF2004} allow to address the physics of time-dependent and out of equilibrium systems.
The crux of the DMRG algorithm is a decimation procedure that chooses the physically most relevant states to describe the target states. It is now known that DMRG  works well because 
the ground states of non-critical quantum chains like the XXZ chain are only slightly entangled, i.e. they obey an area law of entanglement that says 
that the entanglement between a distinguished block of the chain and the rest of the chain is bounded by the boundary area of the block.
In fact the DMRG procedure is a variational ansatz over states known as Matrix product states (MPS) \cite{FNW1992}. The standard DMRG procedure and its connection with
MPS and entanglement is described in detail in \cite{US2005}. 
For a single XXZ chain our target states are the ground state $\psi_0$ and first excited state $\psi_1$ restricted to
a sector of magnetization. 
We use the standard DMRG procedure with the adaptation that we grow the chain while restricting 
the blocks to the sector of zero magnetization using the symmetry of the Hamiltonian (see \cite{MNS2007}).

For the two-qubit gates we convert the two chain system to a one dimensional spin chain by a spin ladder construction.
\begin{equation*}
\begin{array}{ccccccccccc}
\Hil_{-L+1}^{(1)} & \otimes & \Hil_{-L+2}^{(1)} & \otimes 
  & \dots & \otimes & \Hil_L^{(1)} & = & \Hil_{[-L+1,L]}^{(1)}\\
\otimes & & \otimes & & & & \otimes && \otimes \\
\Hil_{-L+1}^{(2)} & \otimes & \Hil_{-L+2}^{(2)} & \otimes 
  & \dots & \otimes & \Hil_L^{(2)} & = & \Hil_{[-L+1,L]}^{(2)}
\end{array}
\end{equation*}
The single site Hilbert space for the DMRG is the rung composed of $\Hil_{\alpha}^{(1)}\otimes\Hil_{\alpha}^{(2)}$ for $\alpha\in[-L+1..L]$. On this site we define the local operators
\begin{equation*}
S^{i(1)}_{\alpha}= S^i_{\alpha}\otimes\unity_{\alpha}, \qquad S^{i(2)}_{\alpha}= \unity_{\alpha}\otimes S^i_{\alpha} \qquad \text{for i= 1,2,3}
\end{equation*}
We can then write the Hamiltonian of this single chain using the above construction
\begin{align*}
H_{L}^{\rm k}(\Delta^{-1})^{(1,2)}    &:= \sum_{\alpha =-L+1}^{L-1}  h^{(1)}_{\alpha,\alpha+1}(\Delta^{-1}) + h^{(2)}_{\alpha,\alpha+1}(\Delta^{-1}) \\
h^{(k)}_{\alpha,\alpha+1}(\Delta^{-1})&=  \Spin^2-S_\alpha^{3(k)} S_{\alpha+1}^{3(k)} - \Delta^{-1}\big(S_\alpha^{1(k)} S_{\alpha+1}^{1(k)}\\
				      &+  S_\alpha^{2(k)} S_{\alpha+1}^{2(k)}\big) + \Spin \sqrt{1-\Delta^{-2}} (S_{\a}^{3(k)} - S_{\a+1}^{3(k)})
\end{align*}
for $k=1,2$. We carry out the DMRG procedure as described in the algorithm with the Hamiltonian $H_{L}^{\rm k}(\Delta^{-1})^{(1,2)}$ but we ensure that we keep both the chains
in the magnetization sector 0 by simultaneously diagonalizing $H_{L}^{\rm k}(\Delta^{-1})^{(1,2)}$ with the total magnetization operators 
\begin{equation*}
S_{tot}^{(k)} 	= \sum_{\alpha=-L+1}^L S^{3(k)}_{\alpha} \qquad \text{for k=1,2}
\end{equation*}
The target states $\psi_0\otimes\psi_0$, $\psi_0\otimes\psi_1$, $\psi_1\otimes\psi_0$, $\psi_1\otimes\psi_1$ are the simultaneous eigenvectors of the these operators and form
the computational basis $\ket{00}$, $\ket{01}$,$\ket{10}$ and $\ket{11}$ for two-qubit quantum computation.

To compute the time evolution of the chain under the controlled evolution by the external fields we use the time-dependent DMRG procedure. The idea is that a two site operator can be applied
to a DMRG state most effectively by expressing the state in the basis where the left block has length $x-1$ so the two middle sites that are untruncated are the the sites where
the operator is acting. We can write the time evolution in the Trotter decomposition
\begin{align*}
e^{-iH\delta} \cong e^{-\frac{i}{2}h_{-L+1,-L+2}}e^{-\frac{i}{2}h_{-L+2,-L+3}}\cdots \\ \cdots e^{-\frac{i}{2}h_{L-2,L-1}}e^{-\frac{i}{2}h_{L-1,L}} +O(\delta^3)
\end{align*}
To apply $e^{-iH\delta}$ to the ground and excited states in the basis with the center sites all the way to the left we apply $e^{-\frac{i}{2}h_{-L+1,-L+2}}$.
After shifting one site to the right we apply $e^{-\frac{i}{2}h_{-L+2,-L+3}}$ etc. Since all our controls are two site controls at the center, only the interaction
$h_{0,1}$ is time-dependent. In the adaptive time-dependent methods the Hilbert space is continuously modified as time progresses by carrying out reduced basis
transformations on the evolved state. In our case since the gates are obtained in a relatively short period of time our Hilbert space remains unchanged resembling the static DMRG methods.


\section{Results} 
\label{sec:Results}
In this section we present numerical results of the construction of quantum gates using the spin-3/2 XXZ spin chain. Our results are for the universal set of single qubit
gates consisting of the Not (X), Hadamard (H), Pi-8 (T) and Phase (S) gates and the two-qubit C-Not gate. All  results are obtained using the DMRG algorithm and the optimal control
methods described in the previous sections. 
\begin{table}
\begin{ruledtabular}
\begin{tabular}{cccc}
Not(X)       & Hadamard(H) & Pi-8(T) & Phase(S)\\\hline
0.1874       &  -0.2182   & -0.1152& -0.0797\\
-0.0533      &  -0.1176   & -0.2544  & -0.1889\\
-0.2447      &  -0.0631   & -0.3310  & -0.2579\\
-0.3587      &  -0.0670   & -0.3613  & -0.2945\\
-0.3764      &  -0.1296   & -0.3632  & -0.3085\\
-0.2901      &  -0.2396   & -0.3524  & -0.3091\\
-0.1075      &  -0.3766   & -0.3410  & -0.3031\\
0.1376       &   -0.5154  & -0.3358  & -0.2943\\
0.3712       &   -0.6286  & -0.3383  & -0.2836\\
0.4908       &  -0.6917   & -0.3443  & -0.2691\\
0.4355       &  -0.6899   & -0.3441  & -0.2466\\
0.2359       &  -0.6241   & -0.3222  & -0.2103\\
-0.0246      & -0.5099    & -0.2590  & -0.1538\\
-0.2681      & -0.3723    & -0.1328  & -0.0715\\
-0.4399      &  -0.2404   & 0.0709   & 0.0386\\
-0.5065      &  -0.1424   & 0.3368   & 0.1704\\
-0.4553      &  -0.1001   & 0.5877   & 0.3053\\
-0.2959      &  -0.1225   & 0.7029   & 0.4128\\
-0.0605      &  -0.2033   & 0.6294   & 0.4642\\
0.1909       &  -0.3236   & 0.4374   & 0.4516\\
\end{tabular}
\end{ruledtabular}
\caption{Numerical simulation of the construction of the Not, Pi-8, Hadamard and  Phase gates. Results are obtained using DMRG and time-dependent DMRG 
for a spin-$\frac{3}{2}$ chain of $L=50$ sites and at $\Delta^{-1}=0.3$ in the sector corresponding to $M=0$.
The table shows the values of the control field $v_1(t)$ with gate time $T=10$ discretized with $\Delta t = 0.5$.}
\label{table:qbitcontrol}
\end{table}
\begin{table}\begin{ruledtabular}
\begin{tabular}{ccc}
    $v_1(t)$  &  $v_2(t)$   &  $v_3(t)$\\\hline
   -0.4040 &   0.3953 &   0.0540\\
    2.4494 &   0.2588 &   0.0501\\
    3.7163 &   0.1886 &  -0.1314\\
    3.0455 &   0.1766 &  -0.2677\\
    1.6565 &   0.2185 &  -0.0872\\
    0.5583 &   0.3085 &   0.1455\\
   -0.1036 &   0.4206 &   0.2346\\
   -0.5117 &   0.5196 &   0.1648\\
   -0.8215 &   0.5716 &  -0.0282\\
   -1.0083 &   0.5578 &  -0.3599\\
   -0.8630 &   0.4926 &  -0.9102\\
   -0.3774 &   0.3612 &  -1.4955\\
   -0.0810 &   0.0978 &  -1.4905\\
   -0.0940 &  -0.1214 &  -0.9196\\
   -0.0881 &  -0.1138 &  -0.3217\\
   -0.0408 &   0.0794 &   0.0922\\
   -0.1805 &   0.3399 &   0.3374\\
   -0.8094 &   0.5868 &   0.4146\\
   -2.2183 &   0.8522 &   0.2021\\
   -4.2425 &   1.2694 &  -0.4760\\
\end{tabular}
\end{ruledtabular}
\caption{The C-Not gate controls using two spin-$\frac{3}{2}$ XXZ-chains of length $L=50$
at $\Delta^{-1} =0.25$ and $M=0$ for both the chains. The gate time $T=3.5$ is discretized into $N=20$ time steps. The table shows the values for the three control fields $v_1$, $v_2$ and $v_3$ that are constant during any one of the time intervals.}
\label{table:cnotcontrol}
\end{table}

\begin{figure}
$$\begin{array}{cc}
\epsfig{file=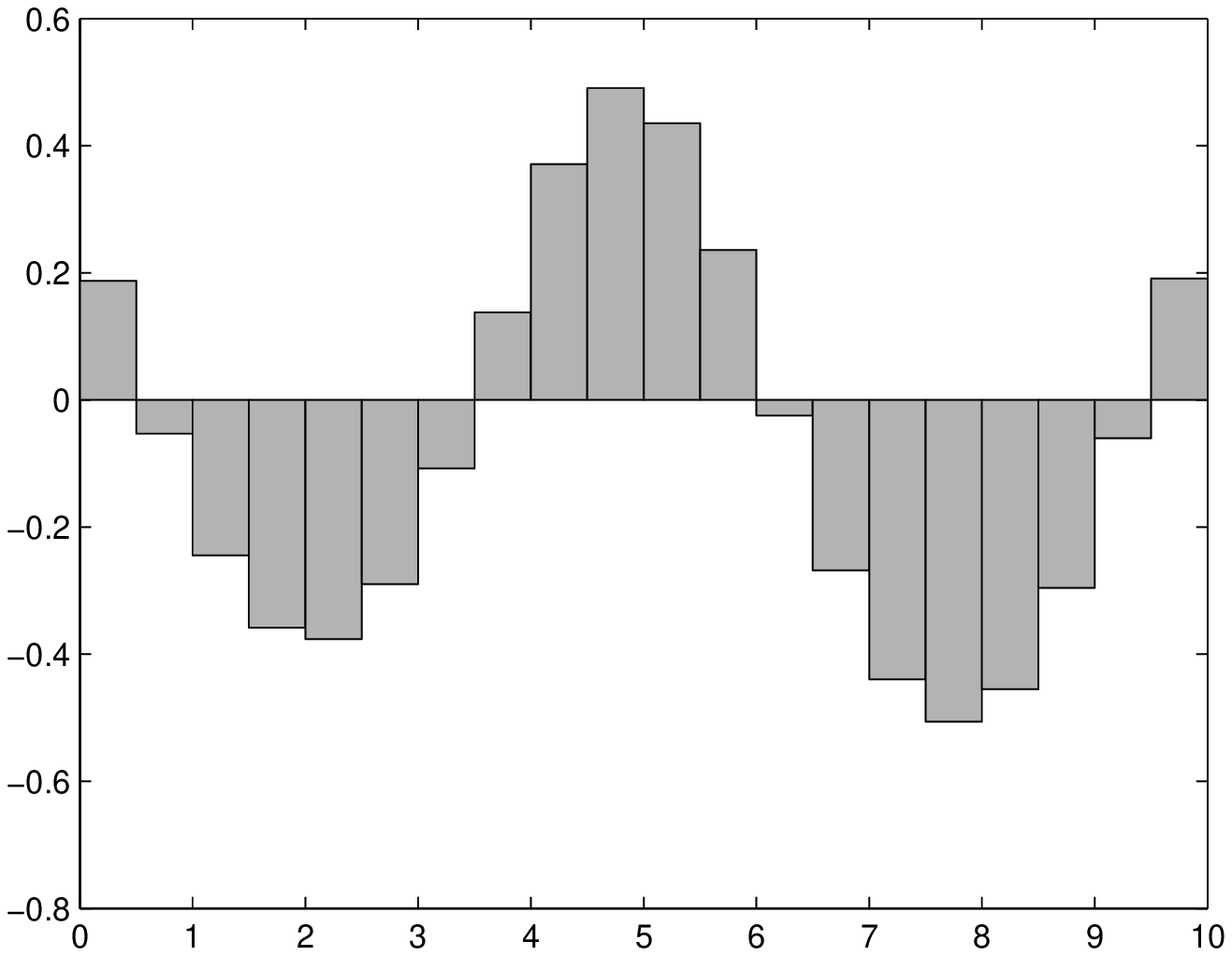,height=3.1cm,width=4cm}&\epsfig{file=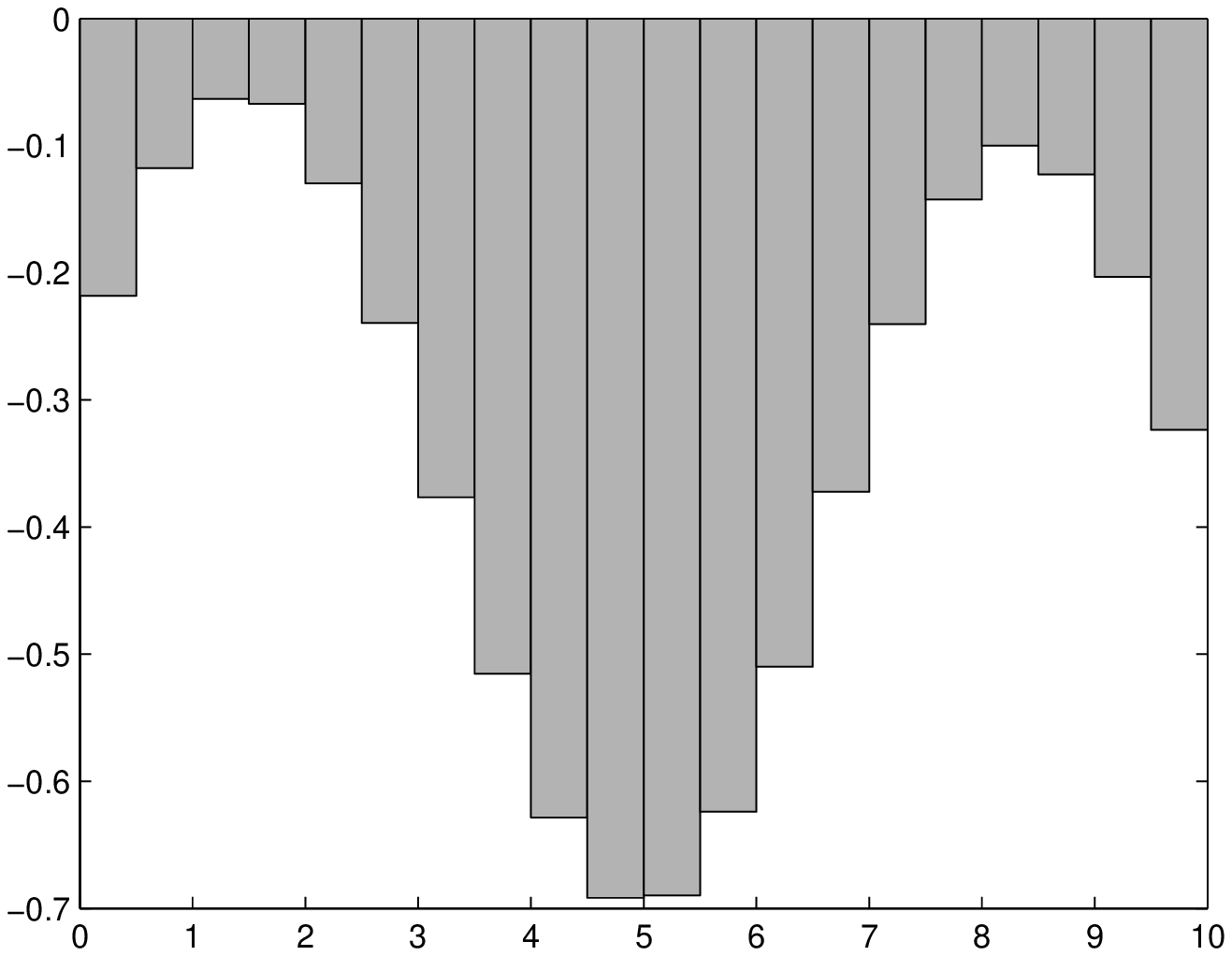,height=3.1cm,width=4cm}\\\\
\epsfig{file=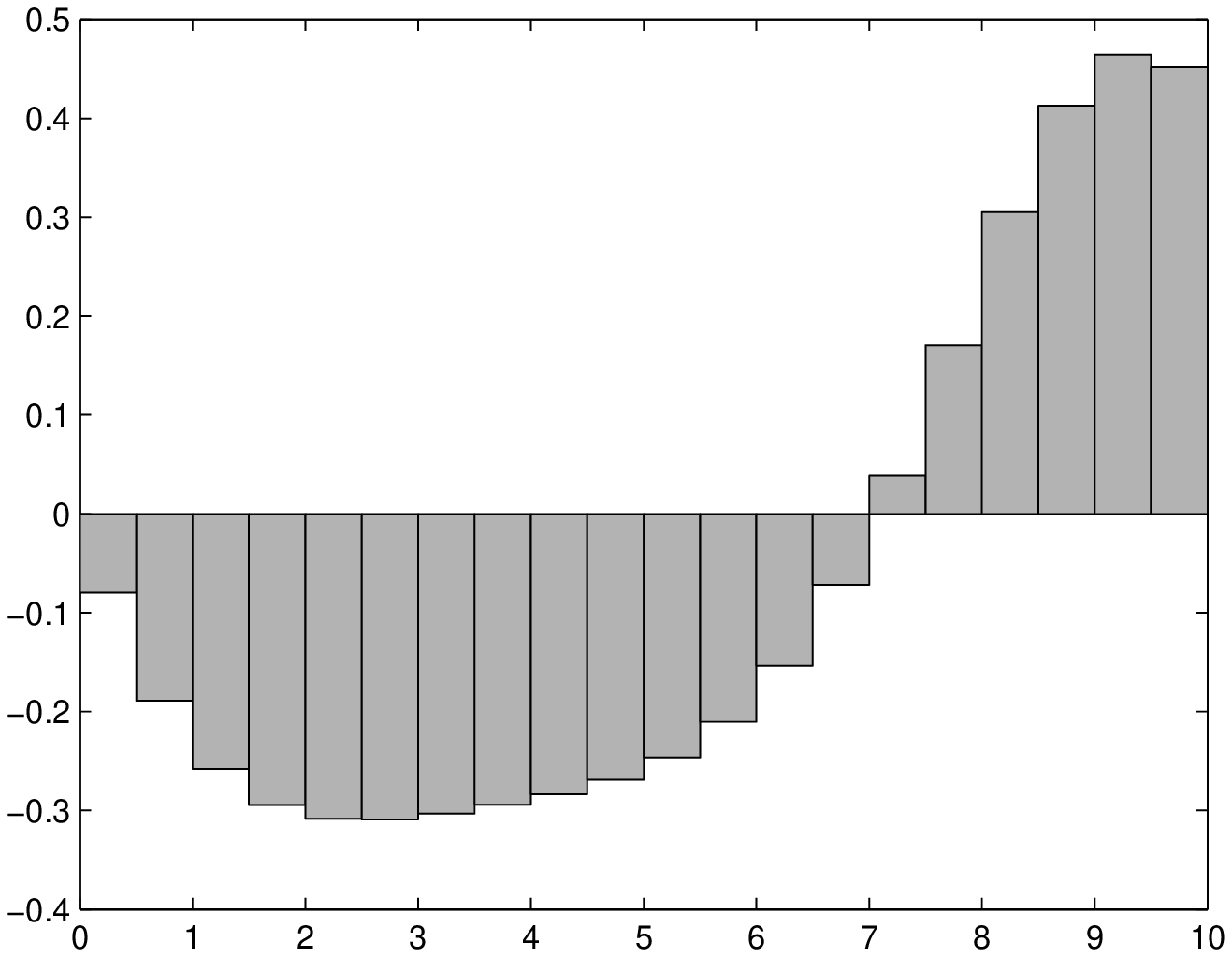,height=3cm,width=4cm}&\epsfig{file=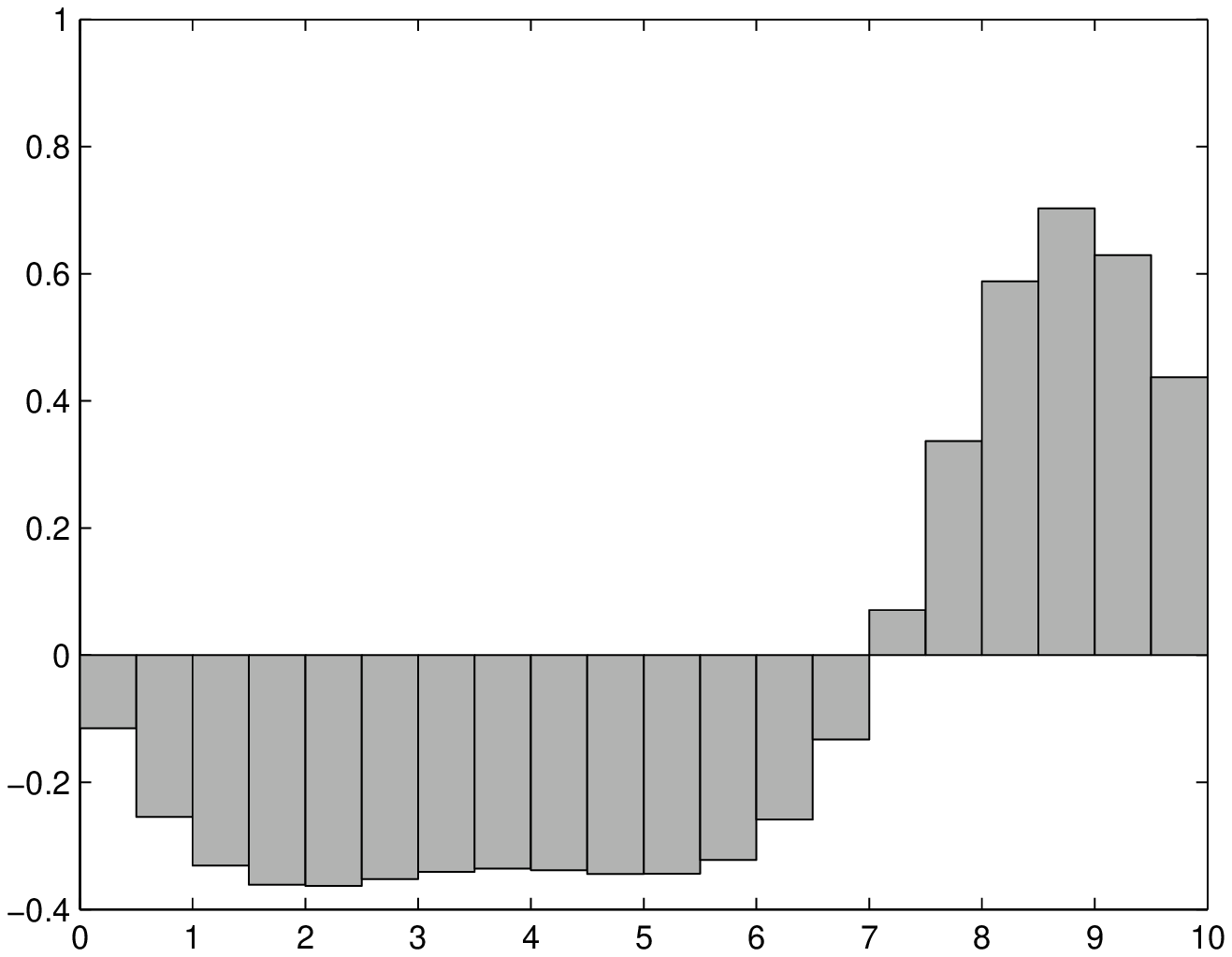,height=3cm,width=4cm}
\end{array}
$$
\caption{(Clockwise) The controls for Not, Hadamard, Pi-8 and Phase gates controls plotted versus time discretized for time steps of $\Delta t = 0.5$. }
\label{fig:controls}
\end{figure}

The steps carried out to obtain the single qubit gates are as follows
\begin{enumerate}
\item
We use ground state DMRG of the XXZ chain to obtain the  lowest eigenvectors $\psi_0$ and $\psi_1$ of $H_L^{\rm k}(\Delta^{-1})$ in the sector corresponding to $M =0$. 
\item
We obtain the projected $2\times 2$ control system of equation (\ref{eqn:control4}) with matrices $H$ and $B_1$ with matrix elements given by (\ref{eqn:matrix2}).
For a  target gate $U_f$ and a suitable final time $T$ we find the optimal control $v_1(t)$ on this $2\times 2$ system using
the technique described in section \ref{sec:optimal control}.
\item
Finally we apply the time-dependent DMRG procedure of section \ref{sec:DMRG} to the chain of (\ref{eqn:control1}) for a specified time $T$ starting from 
$\psi_0$ and $\psi_1$ and using the $v_1(t)$ found in step 2 to get the time evolved states $\psi_0(T) = U(T)\psi_0$ and
$\psi_1(T) = U(T)\psi_1$. We compute the induced evolution on the subspace $\mathcal{D}$ to obtain the gate $U_{xxz}$ given by the matrix elements $\ip{\psi_i}{\psi_j(T)}$ for $i,j=0,1$
and compare the overlap with $U_f$ using equation (\ref{eqn:fidelity}).
\end{enumerate}Our desired single qubit target gates are given by the unitaries
\begin{equation*}
\begin{array}{ll}
X=
\begin{pmatrix}
0&i\\
i&0
\end{pmatrix}&
H=
\frac{1}{\sqrt{2}}\begin{pmatrix}
i&i\\
i&-i
\end{pmatrix}\\\\
T=
\begin{pmatrix}
e^{-i\pi/4}&0\\
0&e^{-i\pi/4}
\end{pmatrix}&
S=
\begin{pmatrix}
e^{-i\pi/8}&0\\
0&e^{-i\pi/8}
\end{pmatrix}
\end{array}
\end{equation*}
The gates obtained using the XXZ chain and their fidelities are as follows.
\begin{eqnarray*}
X_{xxz}&=&
\begin{pmatrix}
0.0016 - 0.0011i&0.0033 + 0.9997i\\
-0.0017 + 0.9997i&0.0017 + 0.0011i
\end{pmatrix}\\
\mathcal{F}_{X} &=& 0.9997,\\\\
H_{xxz}&=&
\begin{pmatrix}
-0.0027 + 0.7081i& 0.0011 + 0.7053i\\
-0.0016 + 0.7052i&-0.0022 - 0.7085i
\end{pmatrix}\\
\mathcal{F}_{H} &=& 0.9995,\\\\
T_{xxz}&=&
\begin{pmatrix}
0.9221 - 0.3859i&-0.0037 + 0.0038i\\
0.0037 + 0.0038i&0.9216 + 0.3871i
\end{pmatrix}\\
\mathcal{F}_{T} &=& 0.9995,\\\\
S_{xxz}&=&
\begin{pmatrix}
0.7043 - 0.7095i & -0.0046 + 0.0015i\\
0.0045 + 0.0016i&0.7017 + 0.7121i
\end{pmatrix}\\
\mathcal{F}_{S} &=& 0.9997
\end{eqnarray*}
The optimal controls $v_1(t)$ used to get the gate results are shown in Table \ref{table:qbitcontrol} and Figure \ref{fig:controls}.
For the C-Not gate the procedure described earlier is only slightly modified. 
We do the ground state DMRG of a one dimensional chain built from the spin ladder described in section \ref{sec:DMRG}
to get four eigenvectors $\psi_{mn}$ for $m,n = 0,1$.
The optimal control procedure is applied to the $4\times 4$ control system (\ref{eqn:control4}) with $H$, $B_1$, $B_2$, $B_3$ given by equations 
(\ref{eqn:matrix4}) to find the controls $v_1(t)$, $v_2(t)$ and $v_3(t)$.
The time-dependent DMRG procedure is applied to the  chain of equation (\ref{eqn:control3}) for time $T$ with the controls $v_1(t)$, $v_2(t)$ and $v_3(t)$
to get the time evolved states $\psi_{mn}(T) = U(T)\psi_{mn}$. The induced evolution on the subspace $ \mathcal{D}_1\otimes\mathcal{D}_2$ gives the ${\rm C-Not}_{xxz}$ gate with matrix elements $\ip{\psi_{mn}}{\psi_{rs}(T)}$.
Table \ref{table:cnotcontrol} shows the optimal controls $v_1(t)$, $v_2(t)$ and $v_3(t)$ used to obtain the C-not gate.
The gate obtained using the $XXZ$ chain and gate fidelity is as follows
\begin{widetext}
\begin{eqnarray*}
\rm {C-Not}&=&
\begin{pmatrix}
1&0&0&0\\
0&1&0&0\\
0&0&0&1\\
0&0&1&0
\end{pmatrix}\\\\
\rm {C-Not}_{xxz} &=& 
\begin{pmatrix}
0.9959+0.0001i&-0.0015+0.0006i&0.0003-0.0003i&-0.0010-0.0001i\\\\
-0.0014-0.0010i&0.9939-0.0003i&0.0015+0.0000i&-0.0005+0.0005i\\\\
0.0013-0.0001i&0.0004+0.0003i&0.0004-0.0003i &0.9945-0.0008i\\\\
-0.0003-0.0003i&-0.0013+0.0002i&0.9954-0.0004i&0.0003+0.0003i
\end{pmatrix}\\\\
\mathcal{F}_{\rm{C-Not}} &=& 0.9949\\
\end{eqnarray*}
\end{widetext}


\begin{acknowledgments}
Based upon work supported in part by the National Science
Foundation under Grants DMS-0605342 and DMS-0757581. 
J.M. also received support from NSF VIGRE grant DMS-0636297.
\end{acknowledgments}

\end{document}